\documentclass[10pt,conference]{IEEEtran}
\IEEEoverridecommandlockouts
\usepackage{cite}
\usepackage{amsmath,amssymb,amsfonts}
\usepackage{graphicx}
\usepackage{textcomp}
\usepackage{xcolor}
\usepackage{booktabs}
\usepackage{multirow}
\usepackage{algorithm}
\usepackage{algpseudocode}
\usepackage{enumitem}
\usepackage{url}
\usepackage{tabularx}
\usepackage{multirow}
\usepackage{balance}
\usepackage{cite}   %

\usepackage{color}

\usepackage{xspace}

\usepackage{enumitem}

\usepackage[most]{tcolorbox}
\newcounter{findingCounter}

\newcommand{\finding}[1]{
  \begin{tcolorbox}[enhanced, left=3mm,right=3mm,
    colback=gray!10, colframe=gray!80, boxrule=0pt,
    borderline west={4pt}{0pt}{gray!90},
    breakable
    ]
    \stepcounter{findingCounter}
    \textbf{Finding \thefindingCounter:} #1
    \end{tcolorbox}
}

\def\BibTeX{{\rm B\kern-.05em{\sc i\kern-.025em b}\kern-.08em
    T\kern-.1667em\lower.7ex\hbox{E}\kern-.125emX}}

\begin{document}

\title{Does It Render Everywhere? A Study of Cross-Environment Compatibility in MLLM-Generated Webpages}

\author{Ziyun Guo$^1$, Jingyu Xiao$^2$, Yuqiang Sun$^3$, Yintong Huo$^1$ \thanks{Yintong Huo is the corresponding author.}\\
$^1$Singapore Management University, Singapore\\
$^2$The Chinese University of Hong Kong, Hong Kong SAR, China\\
$^3$Nanyang Technological University, Singapore}

\maketitle

\begin{abstract}

Multimodal Large Language Models (MLLMs) have been increasingly adopted to automate webpage generation from visual designs (e.g., screenshots). However, existing evaluations are limited to visual fidelity assessment under a fixed browser-device configuration. Such a setting overlooks the cross-environment rendering compatibility for real-world deployments.

To address this gap, we present the first systematic empirical study of cross-environment compatibility in AI-generated webpages. 
Specifically, we construct \textsc{WebCompat}, a dataset of 2,032 annotated instances, comprising webpages generated by 8 representative AI tools, each rendered across 9 browser-and-device combinations. Based on the dataset, we conduct an analysis on the prevalence of compatibility issues, the user-perceptible visual symptoms they manifest, and their underlying code-level root causes.
Our findings reveal that 68\% of generated webpages exhibit at least one compatibility issue, underscoring the pervasive reliability concerns surrounding MLLM-generated front-end artifacts.
The most prevalent symptoms are failures that disrupt the entire page layout (88.3\%): pages shrink directly to fit the target screen with too small fonts, or exhibit scale mismatches that produce cut-off content. Failures localized to individual elements, such as image distortion or missing components, are comparatively less common (13.4\%). 
Furthermore, although most MLLMs incorporate responsive design patterns into the generation, they fail to properly implement these codes.

Guided by the findings, we develop \textsc{XCompat}, a lightweight offline compatibility issue detector that combines visual screenshots and the structural DOM tree for analysis. It achieves an F1 score of 0.903 on the WebCompat-test, outperforming the existing compatibility checking tools and LLM baselines. All datasets and tools are released to support future research on rendering reliability in MLLM-based front-end code generation.

\end{abstract}
\begin{IEEEkeywords}
Multimodal code generation, UI2Code, evaluation, cross-environment compatibility
\end{IEEEkeywords}

\section{Introduction}

Multimodal large language models (MLLMs) have been widely adopted to produce front-end code from screenshots or natural-language requirements~\cite{wu2025mllm,chen2026designcoder,yun2024web2code,yuan2025designrepair}.
This capability has driven the development of numerous AI-assisted web development products, representing a rapidly expanding market~\cite{vercelV0Screenshots,cursorProduct}. 
Concurrently, academic research has formalized this problem as the UI-to-code (i.e., screenshot-to-code) task, yielding a growing body of benchmarking~\cite{si2025design2code, xiao2024interaction2code, xiao2025designbench} and methodology studies~\cite{wan2025dcgen}. 
Despite competitive benchmark scores, it remains unknown whether MLLM-generated front-end code is ready for real-world deployment, where it must render correctly across diverse browsers and devices~\cite{wei2016taming,huang2023conffix,zhao2020seenomaly,liu2020owl}.

\begin{figure}[t]
  \centering
  \includegraphics[width=\columnwidth]{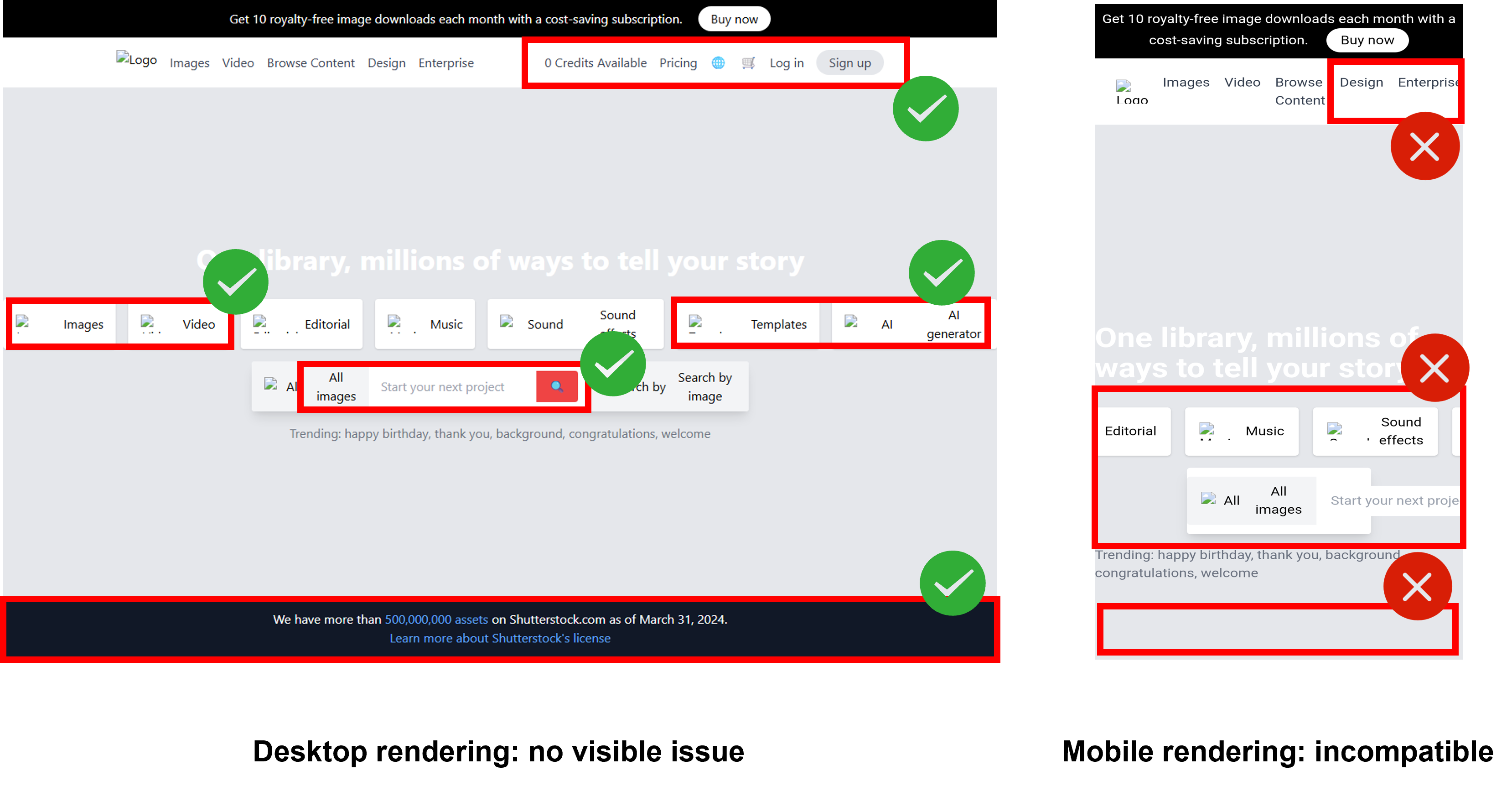}
  \caption{An example showing that correct rendering in one environment does not guarantee cross-environment compatibility. The generated webpage appears correct in the desktop environment, but exhibits element overflow and visibility issue when rendered in a mobile environment.}
  \label{fig:intro_example}
\end{figure}

Existing UI-to-code work~\cite{si2025design2code,wan2025dcgen,chen2018ui,xiao2024interaction2code} emphasizes the generation quality through visual fidelity.
Given a visual design (screenshot) as input, UI-to-code asks models to generate the front-end code that could reproduce the design.
Current benchmarks then compare an AI-generated webpage with its reference under \textit{a fixed rendering setup}, using metrics such as image similarity, text matching, or code-level similarity~\cite{si2025design2code,wan2025dcgen,xiao2025designbench}. 
However, real-world webpages are accessed across different \textit{environments}, spanning various devices and browsers.
As a result, a page that renders correctly under one configuration may still fail in other environments, manifesting as layout shifts, content overflow, or visual distortion. Figure~\ref{fig:intro_example} illustrates this gap: the generated page renders correctly in desktop rendering, but exposes an element overflow and visibility issue in a mobile environment.
This raises a natural question: \textbf{do AI-generated webpages suffer from cross-environment rendering compatibility issues?}

To answer the question, we conducted the first empirical study on compatibility issues in UI-to-code generation. 
We defined an incompatibility as a \textit{user-perceivable visual failure} that is not attributable to intended responsive adaptation and that degrades the webpage’s presentation, content accessibility, or usability. %
Specifically, we propose \textsc{WebCompat}, a dataset covering webpages from representative AI tools across three categories: commercial UI-to-code products, frontier MLLMs, and academic open-source systems.
We rendered each webpage across 9 environments spanning multiple browsers and devices, yielding 2,032 instances for compatibility analysis.  
We then examined the prevalence of compatibility issues in these webpages, characterize their visual symptoms, and uncover the underlying causes at the code level. 

Our study finds that:
\textbf{(1)} Incompatibility is pervasive across AI-generated webpages. Across the eight generators, 68\% of generated webpages contain at least one cross-environment compatibility issue on average.
\textbf{(2)} AI-generated webpages fail more often than human-written webpages (40\%) under the same rendering settings.
\textbf{(3)} Cross-device incompatibility dominates cross-browser incompatibility. Among generated webpages, 67.5\% exhibit cross-device-only failures, while only 1.0\% exhibit cross-browser-only failures.
\textbf{(4)} The observed failures are mostly page-level. Page-level symptoms account for 88.3\% of cases and are dominated by direct page shrink, initial scale mismatch, and bottom whitespace with viewport adaptation issues.
\textbf{(5)} While the AI-generated pages often include responsive layout design, they still fail due to incomplete responsive implementations.The dominant causes are missing viewport meta (46\%) and missing flexible wrapping (29.5\%)

Based on empirical findings, we observe that existing compatibility testing tools cover only part of the incompatibilities in MLLM-generated webpages. To automate the comprehensive detection, we design \textsc{XCompat}, a lightweight detector that analyzes screenshots and post-render DOM trees to identify the aforementioned incompatibilities. On the test set, \textsc{XCompat} achieves an F1 score of 0.903 and an accuracy of 0.953 over 408 instances, outperforming all baselines while maintaining high efficiency.

In summary, this paper makes the following contributions.
\begin{itemize}[leftmargin=*]
    \item \textbf{Problem formulation.} We introduce cross-environment compatibility as a quality dimension for AI-generated webpages and formulate it as a rendered output comparison problem.
    \item \textbf{Benchmark.} We construct \textsc{WebCompat}, a dataset with 2{,}032 annotated instances of AI-generated webpages from multiple generators and rendering environments.
    \item \textbf{Empirical findings.} We show that compatibility issues are pervasive in AI-generated webpages and further derive five findings on their failure symptoms and causes.
    \item \textbf{Detection method.} We develop \textsc{XCompat}, a lightweight offline detector that outperforms existing compatibility testing tools and multimodal LLM baselines.
\end{itemize}

\section{Related Work}

\subsection{Compatibility Testing for Web Applications}
Existing research on web compatibility testing has focused on three directions: cross-browser incompatibility (XBI), cross-device environments (XDI), and responsive layout failures. The earliest work~\cite{mesbah2011automated} characterized XBI as a pervasive, costly, and understudied defect in web engineering. A line of following work utilizes differential testing to compare DOM structures, visual renderings, and behavioral states across different rendering engines, as demonstrated by tools like WebDiff~\cite{choudhary2010webdiff}, X-PERT~\cite{choudhary2013x}, and CrossCheck~\cite{crosscheck}. Complementarily, Browserbite~\cite{saar2016browserbite} approaches XBI purely through screenshot comparison, using image features and machine learning to predict incompatibilities without requiring DOM access. The differential paradigm extends naturally to XDI, where device-cloud platforms such as BrowserStack execute visual regressions across a matrix of browser, hardware, and OS configurations. Since it involves more drastic viewport changes, most practices still rely on manual inspection after simulated rendering. More recently, ReDeCheck~\cite{redecheck} employs an oracle-free, graph-based analysis to enumerate viewport widths and detect failures in element alignments, flagging cases where the front-end code breaks down.

While prior web compatibility research exclusively evaluates human-authored code, the unique cross-environment failures of AI-generated webpages remain entirely unexplored. Furthermore, existing detection tools target narrow, individual defect classes (such as isolated browser or viewport issues). Our work establishes the first unified benchmark and holistic detection framework for compatibility failures produced by AI coding.

\subsection{UI-to-Code Methods and Evaluations}
Translating visual designs into code is challenging and time-consuming, as it requires domain knowledge to map GUI components to spatial layouts. MLLMs have demonstrated strong capability in this task. 
Benchmarking studies such as Design2Code~\cite{si2025design2code} evaluate models on generating HTML/CSS implementations from webpage screenshots, while generation methods such as DCGen~\cite{wan2025dcgen}, UICopilot~\cite{gui2025uicopilot}, and WAFFLE~\cite{liang2025waffle} improve UI-to-code generation through decomposition, hierarchical generation, or structure-aware fine-tuning, respectively. 
Subsequent work has further advanced MLLMs' UI-to-code ability through techniques such as layout guidance~\cite{wu2025mllm}, token compression~\cite{xiao2025efficientuicoder}, component-based implementation~\cite{xiao2026comuicoder} and test-driven development~\cite{wan2026runnable}.

However, existing evaluations focus exclusively on reproduction similarity within a fixed rendering environment. Their metrics, such as visual similarity (SSIM), and code similarity (BLEU), measure reproduction fidelity but fail to reflect the cross-environment reliability of generated webpages.
Neglecting this dimension poses practical risks when deploying AI-generated pages to real-world scenarios.

\section{Dataset Construction}
\label{sec:dataset}

We construct \textsc{WebCompat}, a dataset for analyzing cross-environment compatibility issues of AI-generated front-end code. It consists of webpages produced by eight representative generation tools, rendered on nine devices and browser environments through BrowserStack~\cite{browserstack2026}, with ground-truth compatibility labels from human annotation.

\begin{figure*}[th]
  \centering
  \includegraphics[width=0.9\linewidth]{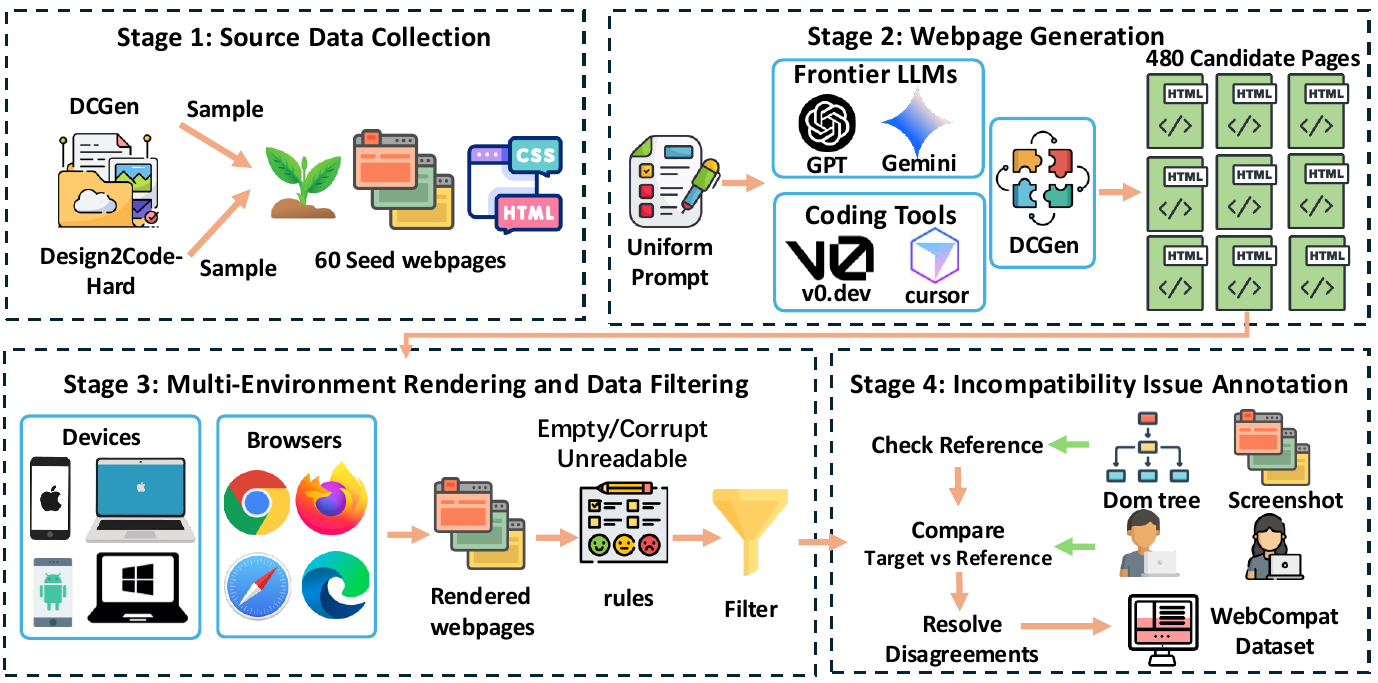}
  \caption{Overview of the WebCompat construction pipeline.}
  \label{BenchmarkPipline}
\end{figure*}

\subsection{Problem Formulation}

We formulate the compatibility issues detection as a pairwise comparison task. 
For an AI-generated page $p$, we render it in two environments, namely reference $e_r$ and test $e_t$. Each rendering yields a screenshot and its post-render DOM, written $R(p,e)=\langle I(p,e), D(p,e)\rangle$.
Given the two renderings, the task $f$ is to decide whether the target has a visual difference relative to the reference:
\begin{equation}
f\big(R(p,e_r),\,R(p,e_t)\big) \to bool(isCompatible)
\end{equation}

\label{sec:task}

\subsection{Construction Pipeline}
\label{sec:pipeline}
We aim to gather a comprehensive dataset capturing the landscape of front-end design automation tools and reveal their performance in multi-environment rendering scenarios.
To this end, we constructed a dataset through a four-stage pipeline, as illustrated in Figure~\ref{BenchmarkPipline}:
(1)~source data collection,
(2)~webpage generation,
(3)~multi-environment rendering, and
(4)~manual annotation on incompatibility.
Each stage is described in detail below.

\subsubsection{Stage 1: Source Data Collection}
\label{sec:stage1}
Our study focuses on single-page instances from established UI-to-code benchmarks, where each page is self-contained in a single file and rendered across web browsers on different devices. We choose single-page static applications because they have stable and deterministic rendering, regardless of user interactions or backend state. This isolates cross-environment rendering issues and allows us to measure compatibility independently.

In particular, we source webpages from two representative datasets, DCGen~\cite{wan2025dcgen} and Design2Code-Hard~\cite{si2025design2code}, both of which provide paired visual inputs and the associated HTML/CSS implementations.
The two datasets differ in both design intention and task difficulty: DCGen is built from real-world websites, while Design2Code-Hard is a harder dataset that deliberately incorporates difficult cases from GitHub Pages, including long pages, intricate HTML structures, and non-English content, which require models to handle more difficult visual-text recognition and layout reproduction.
We randomly sample 30 pages from each, resulting in 60 source pages for the subsequent generation stage.

\subsubsection{Stage 2: Webpage Generation}
\label{sec:stage2}
To reflect current practice in automated front-end development, we employ a diverse set of leading UI-to-code methods to reproduce the source pages, spanning frontier LLMs, commercial agentic coding tools, and academic representatives.
Concretely, we include three frontier LLMs (GPT-4~\cite{achiam2023gpt}, GPT-5.1~\cite{openai2025gpt51}, and Gemini-3-Flash~\cite{googleGeminiCanvas}), two agentic coding tools (v0.dev~\cite{vercelV0Screenshots} and Cursor\cite{cursorProduct}), and three variants of DCGen that use different LLM backends (DCGen-GPT4, DCGen-GPT5.1, and DCGen-Gemini-3-Flash).  DCGen is a leading academic tool based on the current leaderboard~\cite{le2026uibenchkit}.
To ensure a fair comparison, we standardize both prompt design and tool configuration. For DCGen, we follow its original implementation. 
For all other prompt-based generators, we use the same prompt template per instance (adopted from DCGen's direct-prompting variant).
For agentic coding tools, we keep their default configurations to reflect their standard usage conditions. 
This process yields $60\times8=480$ candidate webpages for the next stage.

\subsubsection{Stage 3: Multi-Environment Rendering}
\label{sec:stage3}

This stage renders each generated page in multiple environments to expose its compatibility issues.
To emulate a variety of real-world webpage access scenarios, we consider two dimensions: cross-browser and cross-device.
First, different browsers (e.g., Google Chrome, Firefox, Safari) use different layout calculations, spacing rules, and overflow handling in the rendering engine, all of which may lead to visual inconsistencies.
Second, different devices have different viewport sizes, where a \textbf{viewport} refers to the visible area of a webpage on a screen. 
A webpage that displays correctly on a desktop browser may fail on a mobile device due to its smaller viewport. This can lead to layout issues like element overlap or clipping.
Therefore, we include cross-device environments to capture failures that depend on viewport differences.

Following this design, we render each webpage across nine device-browser environments on BrowserStack~\cite{browserstack2026}, a cloud platform that runs real devices and browsers. 
Table~\ref{tab:device-matrix} lists the corresponding operating systems and browser configurations.

After rendering, we remove pages that are not suitable for cross-environment comparison.
These include pages with blank content, malformed HTML output, failures during rendering and capturing.
We also remove layouts that depend on unsupported interactive behavior such as collapsible menus.
This filtering retains 254 valid webpages.
Then, we set Chrome-on-Windows as the reference environment, because Chrome's Blink engine represents the most widely deployed browser, and desktop applications remain the dominant platform in existing GUI research~\cite{statcounter_browser_2026,statcounter_desktop_os_2026,choudhary2010webdiff,redecheck}.
For each reference, we pair it with each of the other eight renderings as target environments.
This produces $254\times8=2{,}032$ cross-environment comparison pairs for annotation.
\begin{table}[t]
\centering
\footnotesize
\setlength{\tabcolsep}{4pt}
\caption{Device Matrix (BrowserStack)}
\label{tab:device-matrix}
\begin{tabular}{lllll}
\toprule
\textbf{Device} & \textbf{OS} & \textbf{OS Ver.} & \textbf{Browser} & \textbf{Ver.} \\
\midrule
\multicolumn{5}{l}{\textit{Mobile}} \\
\midrule
Google Pixel 7 & Android & 14       & Chrome  & 146.0 \\
iPhone 13      & iOS     & 18       & Safari  & 15.6 \\
iPhone 13      & iOS     & 18       & Chrome  & 146.0 \\
\midrule
\multicolumn{5}{l}{\textit{Desktop --- Viewport 1280$\times$800, Screen 1920$\times$1080}} \\
\midrule
Desktop & Windows & 11       & Edge    & 145.0 \\
Desktop & Windows & 11       & Chrome  & 146.0 \\
Desktop & Windows & 11       & Firefox & 148.0 \\
Desktop & OS X    & Monterey & Chrome  & 146.0 \\
Desktop & OS X    & Monterey & Firefox & 148.0 \\
Desktop & OS X    & Monterey & Safari  & 15.6   \\
\bottomrule
\end{tabular}
\end{table}

\subsubsection{Stage 4: Incompatibility Issue Annotation}
\label{sec:stage4}

Annotation is conducted by two annotators with at least four years of experience in front-end web development and GUI testing.
Prior to the formal annotation phase, they complete a pilot labeling session on a sampled set to align their understanding of compatibility issues.
In the main phase, each rendering pair was first labeled independently by the two annotators following the procedure in Algorithm~\ref{alg:annotate}. The independent annotations achieve a Cohen's kappa of 0.9506, indicating a high level of inter-annotator agreement.
Then, disagreements between the two annotators are resolved through a third adjudication round conducted through consensus discussion.
All disagreements are resolved at the end of the discussion.
\begin{algorithm}[t]
\footnotesize
\caption{Cross-environment compatibility annotation}
\label{alg:annotate}
\begin{algorithmic}[1]
\Require Renderings $R(p,e_r)$ and $R(p,e_t)$
\Ensure \textsc{Compatible} or \textsc{Incompatible}
\If{$e_r$ and $e_t$ share the same device}
  \State compare elements across $D(p,e_r)$ and $D(p,e_t)$
         by position, size, and appearance
  \If{any element differs} \Return \textsc{Incompatible} \EndIf
\Else
  \If{the layout in $R(p,e_t)$ reflows relative to $R(p,e_r)$}
    \If{any element is missing, overflows the viewport, or overlaps another}
      \Return \textsc{Incompatible}
    \EndIf
  \Else
    \If{the scaled layout is distorted} \Return \textsc{Incompatible} \EndIf
  \EndIf
\EndIf
\State \Return \textsc{Compatible}
\end{algorithmic}
\end{algorithm}

\subsection{Dataset Summary}

\begin{figure}[t]
\centering
\includegraphics[width=\linewidth]{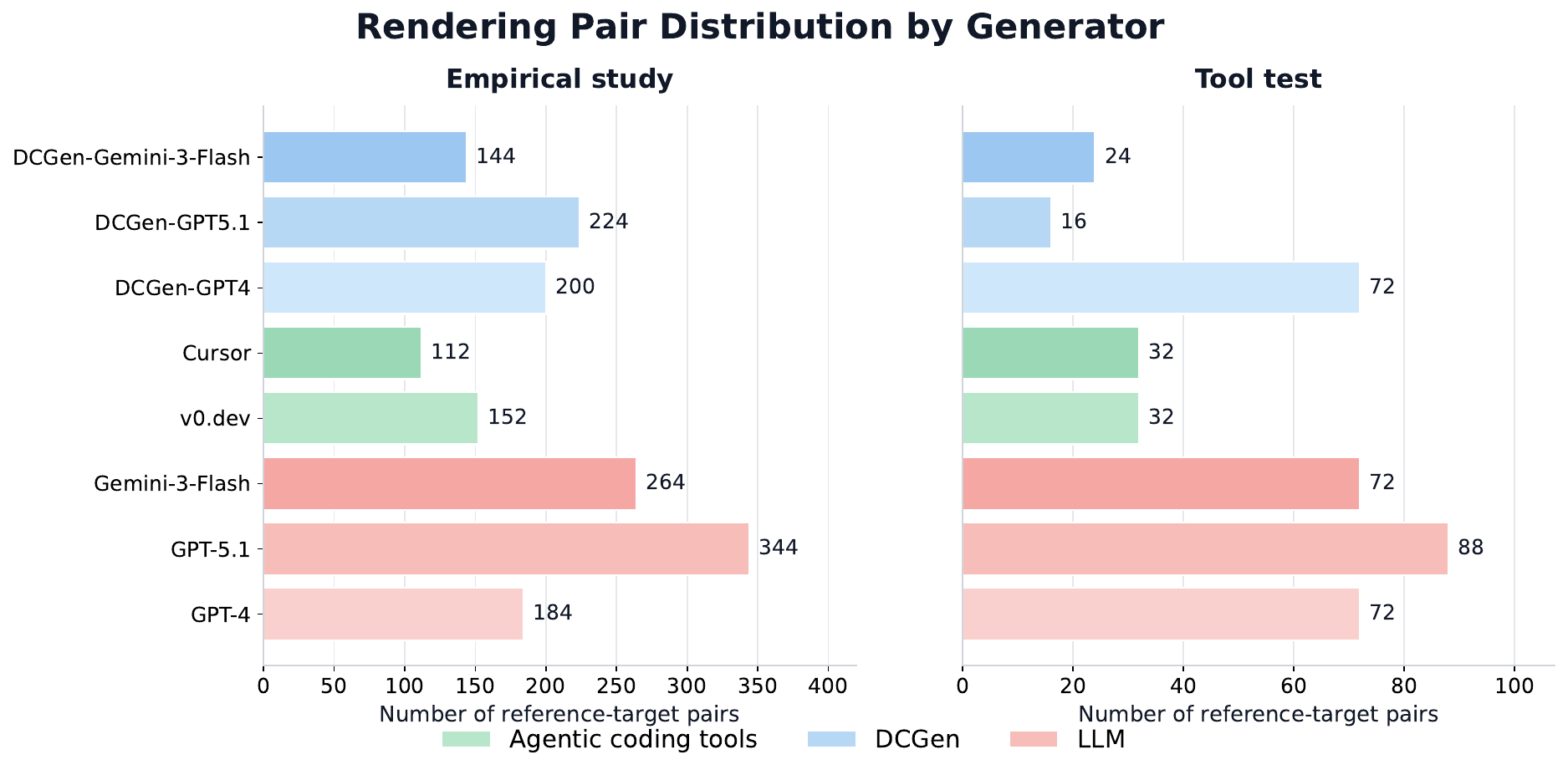}
\caption{Distribution of rendering pairs across the various generators.}
\label{fig:generator_distribution}
\end{figure}

Our dataset comprises 254 valid webpage generations and 2,032 annotated instances (i.e., rendering pairs) across multiple environments. 
Figure~\ref{fig:generator_distribution} shows the distribution of instances across the AI coding tools.
Together, the diversity of websites, AI coding tools, and rendering environments provides the foundation for our study of compatibility issues.

We partition the data 8:2 (i.e., 203:51) at the page level. 
The larger split is used for empirical analysis, while the held-out 20\% serves as a test set for evaluating automated compatibility issue detection to avoid data leakage.

 \section{Empirical Characterization}
\label{sec:empirical}
Building on the dataset described in Section~\ref{sec:dataset}, we structure our empirical study around the following research questions:
\begin{itemize}[leftmargin=*]
        \item \textbf{(RQ1)} \textit{How prevalent are compatibility issues in AI-generated webpages?} We measure generation fidelity and compatibility rates, and compare incompatibility prevalence against human-authored counterparts.
    \item \textbf{(RQ2)} \textit{What symptoms do these compatibility issues manifest?} We manually analyze compatibility issues and derive a taxonomy of cross-environment failure symptoms.
    \item \textbf{(RQ3)} \textit{What are the root causes of these compatibility issues?} We perform code-level analysis to identify sources of rendering inconsistency, yielding insights into subpar coding patterns through MLLM generation.
\end{itemize}

\begin{table*}[t]
\centering
\caption{Generation quality of each generator on the empirical study set. Best score in each row is \textbf{bold}.
\textit{Overall UI Quality} has a maximum score of 15.The Average column reports a sample-weighted mean, where the \textbf{weights are the numbers of rendering pairs} produced by each generator.}
\label{tab:d2c}
\footnotesize
\setlength{\tabcolsep}{2pt}
\renewcommand{\arraystretch}{1.1}
\centering
\begin{tabular}{lccccccccc}
\toprule
& \multicolumn{2}{c}{\textbf{Agent}} 
& \multicolumn{3}{c}{\textbf{DCGen}} 
& \multicolumn{3}{c}{\textbf{Direct}} 
& \\
\cmidrule(lr){2-3} \cmidrule(lr){4-6} \cmidrule(lr){7-9}
Metric & \textbf{Cursor} & \textbf{v0} & \textbf{DCGen-Gemini3} & \textbf{DCGen-GPT4} & \textbf{DCGen-GPT5} & \textbf{Direct-Gemini3} & \textbf{Direct-GPT4} & \textbf{Direct-GPT5} & \textbf{Average}\\
\midrule
\multicolumn{9}{l}{\textit{Visual / Structural Similarity}}\\ 
\midrule 
CLIP & 0.8955 & 0.8715 & \textbf{0.9030} & 0.8441 & 0.8997 & 0.8738 & 0.8204 & 0.8947 &0.8760 \\
SSIM & 0.7807 & 0.7889 & 0.7780 & 0.7899 & \textbf{0.8138} & 0.7994 & 0.7807 & 0.7870 & 0.7913\\
BLEU & 0.6157 & \textbf{0.7360} & 0.7112 & 0.5463 & 0.7290 & 0.7178 & 0.3811 & 0.6941 &0.6491\\
Structure Code Similarity & 0.0597 & \textbf{0.1370} & 0.0812 & 0.1012 & 0.0814 & 0.0916 & 0.1162 & 0.0796 & 0.0928\\
\addlinespace[2pt]
\midrule
\multicolumn{9}{l}{\textit{Fine-grained Visual}} \\
\midrule
BlockMatch & 0.3317 & 0.7553 & \textbf{0.7676} & 0.6821 & 0.6249 & 0.5606 & 0.6666 & 0.7053 & 0.6479\\
Text & 0.6736 & 0.9068 & 0.8752 & \textbf{0.9719} & 0.8278 & 0.8958 & 0.9581 & 0.9406 & 0.8962 \\
Position & 0.5792 & 0.8155 & 0.7821 & \textbf{0.8477} & 0.7568 & 0.7906 & 0.7811 & 0.8386 & 0.7890 \\
Color & 0.6161 & 0.8352 & 0.8117 & 0.8514 & 0.7817 & 0.7918 & 0.8530 & \textbf{0.8742} & 0.8159\\
\addlinespace[2pt]
\midrule
\multicolumn{9}{l}{\textit{MLLM-as-a-Judge }} \\
\midrule
Overall UI Quality  & 7.286 & 6.895 & 9.167 & 6.360 & 7.714 & \textbf{9.424} & 6.174 & 8.558 & 7.852\\
\addlinespace[1pt]
\midrule
\textbf{Incompatibility Rate}  & 0.79 & \textbf{0.26} & 0.50 & 0.84 & 0.93 & 0.42 & 0.43 & 1.00 & 0.68\\
\bottomrule
\end{tabular}
 \end{table*}
\subsection{RQ1: Prevalence of Compatibility Issues}
\textit{1) The Compatibility issues in AI-generated webpages.}

\textbf{Motivation.} The existing UI2Code evaluation compares the reproduction similarity with the original design in a single environment. 
However, a competitive score under such metrics does not reflect whether the same page remains reliable across different browsers and devices. 
Therefore, we first examine the prevalence of compatibility issues in these AI reproductions.

\textbf{Settings.} 
We study the relationship between UI fidelity scores and incompatibility rates on 203 AI-generated webpages from the empirical-study split.
Following previous UI-to-Code work~\cite{si2025design2code,le2026uibenchkit}, we use multiple UI fidelity metrics to cover different aspects of webpage reproduction quality. We report visual similarity metrics such as CLIP~\cite{radford2021learning} and SSIM~\cite{wang2004image} for appearance, BLEU~\cite{papineni2002bleu} for textual similarity, code similarity~\cite{le2026uibenchkit} for HTML/CSS. We also report fine-grained visual matching metrics~\cite{si2025design2code} for element-level quality, and MLLM-as-a-Judge based on WebDevJudge~\cite{li2025webdevjudge} for overall UI quality assessment.

\textbf{Results.} 
Table~\ref{tab:d2c} shows that incompatibility is pervasive across all generations, despite their high visual fidelity score. 
On average, 68\% of AI-generated webpages contain at least one cross-environment compatibility issue. Across all eight tools, the incompatibility range is from 26\% to 100\%. The best-performing v0 stands at 0.26, likely owing to its engineering focus on deployment-ready code. 
This demonstrates that existing AI-produced webpages cannot reliably guarantee cross-environment compatibility. 
\begin{table}[t]
\centering
\caption{Cross-environment incompatibility rates of AI-generated webpages and the
human-written baseline. Incompatibility rates are computed per webpage, not per rendering pair.}
\label{tab:rq1-failure-rates}
\small
\setlength{\tabcolsep}{5pt}
\resizebox{\columnwidth}{!}{%
\begin{tabular}{@{}l l c c c@{}}
\toprule
\textbf{Category} & \textbf{AI Tools} & \textbf{Design2Code} & \textbf{DCGen} & \textbf{Combined} \\
\midrule
\multirow{2}{*}{Agent}
  & Cursor          & 5/6 (83\%)                  & 6/8 (75\%)                   & 11/14 (79\%) \\
  & v0              & 3/7 (43\%)                  & 2/12 (17\%)         & 5/19 (26\%) \\
\midrule
\multirow{3}{*}{DCGen}
  & Gemini-3-Flash  & 4/8 (50\%)                  & 5/10 (50\%)                  & 9/18 (50\%) \\
  & GPT-4           & 11/12 (92\%)                & 10/13 (77\%)                 & 21/25 (84\%) \\
  & GPT-5           & \underline{14/14 (100\%)}   & 12/14 (86\%)                 & 26/28 (93\%) \\
\midrule
\multirow{3}{*}{Direct}
  & Gemini-3-Flash  & 8/15 (53\%)                 & 6/18 (33\%)         & 14/33 (42\%) \\
  & GPT-4           & 7/11 (64\%)                 & 3/12 (25\%)         & 10/23 (43\%) \\
  & GPT-5           & 22/22 (100\%)   & 21/21 (100\%)    & 43/43 (100\%) \\
\midrule
\multicolumn{2}{l}{\textbf{AI overall}} & 74/95 (78\%) & 65/108 (60\%) & 139/203 \textbf{(68\%)} \\
\midrule
\multicolumn{2}{l}{\textit{Human-written baseline}} & & & 24/60 \textbf{(40\%)} \\
\bottomrule
\end{tabular}%
}
\vspace{2pt}
\footnotesize
\end{table}

Across all generators, stronger performance on visual fidelity metrics does not indicate lower incompatibility rates. For example, DCGen-GPT4 achieve the two highest fine-grained visual scores among all models, yet carry a higher incompatibility rate of 0.84. The weak relationship between visual fidelity scores and cross-environment robustness indicates that current benchmarks are insufficient indicators for deployment quality. Although academic models achieve competitive visual fidelity, this does not extend to cross-environment robustness, whereas the commercial tool v0 strikes the most favorable compatibility-fidelity trade-off. This reveals that cross-environment compatibility is a critical quality gap that AI code generation tools must address for real-world deployment.

\finding{Incompatibility is pervasive in AI-generated webpages (average 68\%). Existing visual fidelity scores show weak correlation with cross-environment robustness.}

\begin{table*}[t]
\centering
\caption{Taxonomy of cross-environment failure symptoms in AI-generated
webpages. Class~A failures disrupt global page layout; Class~B failures are
localized to individual elements. }
\label{tab:taxonomy-by-pair}
\footnotesize
\renewcommand{\arraystretch}{1.2}
\begin{tabularx}{\linewidth}{@{}l l X r@{}}
\toprule
\textbf{Level} & \textbf{Type} & \textbf{Definition} & \textbf{\%} \\
\midrule
\multicolumn{4}{@{}l}{\textbf{Class~A: Page-level failures} \hfill 88.3\%} \\
\addlinespace[2pt]
\multirow{3}{*}{\shortstack[l]{A1 \\ (Viewport adaptation \\ failure)}}
  & (A1.1) Shrink-to-Fit   & Page directly shrinks to fit the target screen, making the content too small to read. & 42.1 \\
  & (A1.2) Initial Scale Mismatch  & Initial page scale does not match the device width, causing content to be cut off horizontally or leaving the viewport width only partially occupied. & 23.1 \\
  & (A1.3) Whitespace Anomaly & Unexpected large blank gap at the page bottom after rendering (absent from reference). & 13.8 \\
\addlinespace[2pt]
\multirow{2}{*}{\shortstack[l]{A2 (Structural layout \\ failure)}}
  & (A2.1) Overflow        & Content extends beyond the container/viewport or gets clipped. & 23.1 \\
  & (A2.2) Overlap         & Elements collide and stack with each other, hiding text or controls. & 5.5 \\
\midrule
\multicolumn{4}{@{}l}{\textbf{Class~B: Element-level failures} \hfill 13.4\%} \\
\addlinespace[2pt]
\multirow{3}{*}{B}
  & B1 Image Size Distortion  & Image/placeholder is stretched relative to its intended dimensions. & 10.8 \\
  & B2 Visibility Failure  & Missing elements in target rendering. & 2.1 \\
  & B3 Appearance Deviation   & Element correctly positioned but missing or altered CSS styles change its visual appearance such as color or styling. & 0.5 \\
\bottomrule
\end{tabularx}
\end{table*}

\textit{2) The comparison between AI-generated webpages and human-authored ones.}

\textbf{Motivation.} To understand how widespread compatibility issues are in practice, we compare MLLM generation results with human-written ones, and analyze the primary sources of compatibility challenges.

\textbf{Settings.} Using the 203 valid AI-generated webpages from the empirical-study split, we compare them with 60 human-authored implementations from Design2Code and Design2Code-HARD~\cite{si2025design2code}, which were created before the recent wave of LLM-based webpage generation. We exclude DCGen source pages from the human baseline because many rely on JavaScript-driven behavior, which is outside our static-page setting.
We render the human-authored pages using the same multi-environment pipeline and pairwise annotation procedure in \S\ref{sec:stage3} and \S\ref{sec:stage4}, respectively.The two annotators achieve a Cohen's kappa of 0.943 on the human pair-level labels.

\textbf{Results.} 
AI-generated webpages fail about 1.7$\times$ as often as human-written webpages. As shown in Table~\ref{tab:rq1-failure-rates}, 139 of 203 AI-generated webpages contain at least one compatibility issue, yielding a failure rate of 68\%. The human-written baseline has 24 incompatible webpages out of 60, yielding a failure rate of 40\%.  A Fisher exact test confirms that this difference is statistically significant, $p<0.001$.This result suggests that cross-environment compatibility is already a common and difficult problem in human web development, while AI-generated webpages are even more likely to suffer from such issues.
This gap also appears across the two source groups in the AI set. Design2Code-HARD-based generations fail on 74 of 95 webpages, while DCGen-based generations fail on 65 of 108 webpages. 

\finding{AI-generated webpages exhibit 1.7$\times$ more cross-environment issues than human-written webpages.}

\begin{figure}[t]
  \centering
  \includegraphics[width=\linewidth]{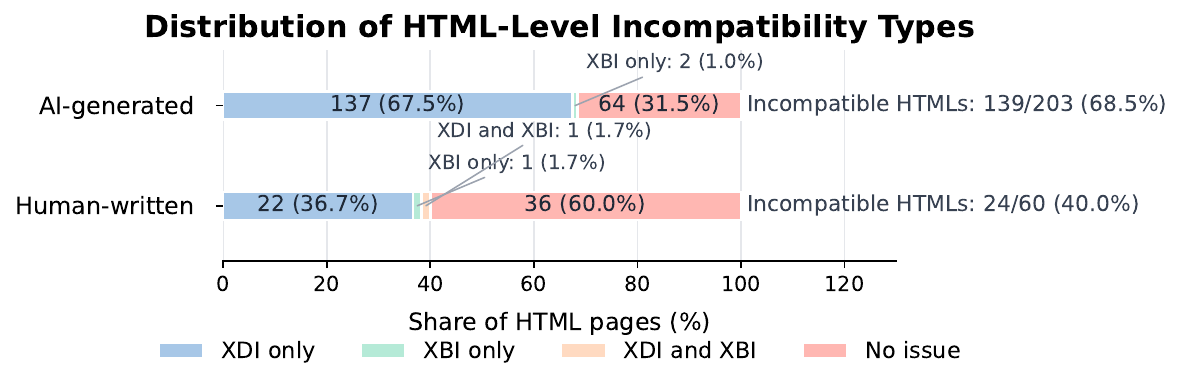}
  \caption{Distribution of XBI, XDI, and issue-free pages for human-written and AI-generated webpages.}
  \label{fig:xbixdi-distribution}
\end{figure}

We further go in-depth into cross-browser incompatibility (XBI) and cross-device incompatibility (XDI), because they stress different aspects of front-end compatibility. 
XBI captures browser-dependent rendering differences under desktop viewport settings. XDI captures whether the same page adapts from desktop to mobile viewport settings.
Figure~\ref{fig:xbixdi-distribution} shows their distribution. XDI (blue bar) is substantially more prevalent than XBI across both groups. Among AI-generated webpages, 67.5\% exhibit XDI-only
failures, while only 1.0\% exhibit XBI-only failures. 
Among human-authored webpages, 36.7\% show
XDI-only failures, compared to just 1.7\% with XBI-only failures and 1.7\% with both. The remaining 36 human-authored pages (60.0\%) are issue-free. These XDI-dominant results indicate the viewport size difference is a greater rendering challenge. While modern browsers have standardized rendering engines, adapting layouts across drastically different screen sizes (e.g., 320px to 1920px) demands comprehensive responsive design logic.

\finding{
Cross-device incompatibility dominates cross-browser incompatibility in both AI-generated and human-written webpages.}

\subsection{RQ2: Failure Symptoms}
\label{sec:taxonomy}
\textbf{Motivation.} While RQ1 shows that most AI-generated webpages have compatibility issues, their binary label does not tell \textit{how} it fails. We therefore study the visual and layout symptoms in incompatible pages to understand their user-perceived impact.

\textbf{Settings.}
Two annotators first inspected screenshots from 20\% of the incompatible webpages as a pilot set. Before the pilot annotation, we provided the annotators with an initial taxonomy adapted from prior work (ReDeCheck~\cite{redecheck} and WebDiff~\cite{choudhary2010webdiff}). During the pilot annotation, they refined this taxonomy into two failure levels (page-level, element-level) and derived an initial set of issue classes. A page-level failure affects the overall layout structure, whereas an element-level failure remains localized to an individual element. Upon this, the two annotators then independently labeled the full set of incompatible cases. When they encountered a new issue class, they discussed it, added it to the guide, and applied it consistently to the remaining cases. Disagreements were collected and resolved through discussion. Before this resolution, the inter-annotator agreement was 0.825, measured by macro-averaged Cohen's $\kappa$ ~\cite{cohen1960coefficient} over all incompatible pairs.

\textbf{Results.} 
Table~\ref{tab:taxonomy-by-pair} presents the failure symptom taxonomy, spanning two class levels (page-level, element-level) with eight failure types, their definitions, and distribution. One instance may exhibit multiple failure symptoms. 
Compatibility failures in AI-generated webpages are predominantly page-level: 88.3\% of incompatible pairs exhibit page-level failures, compared to 13.4\% with element-level failures (note: instances may exhibit multiple failure symptoms). 

\begin{figure}[t]
  \centering
  \includegraphics[width=\linewidth]{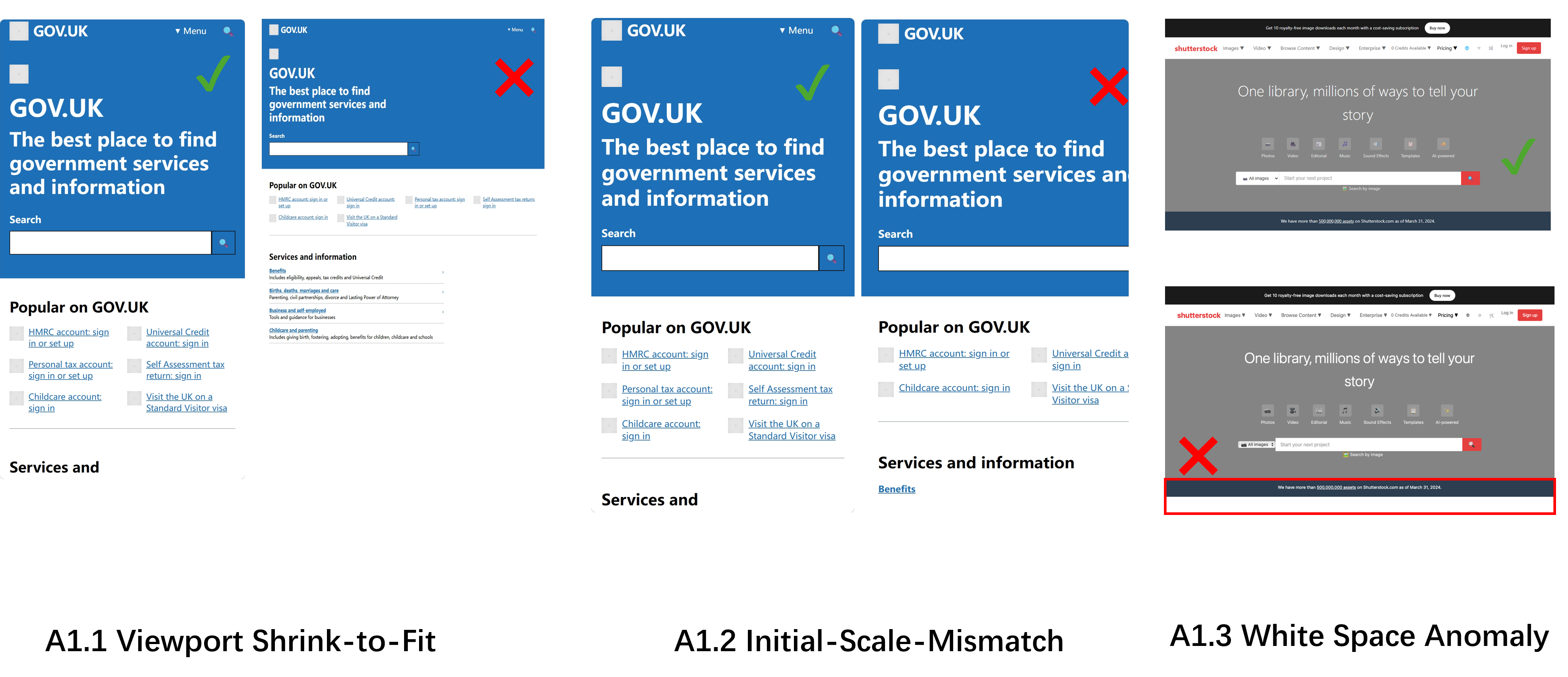}
  \caption{Subclass of A1 (Viewport adaptation failure).}
  \label{fig:subclass-A1}
\end{figure}

\begin{figure}[t]
  \centering
  \includegraphics[width=\linewidth]{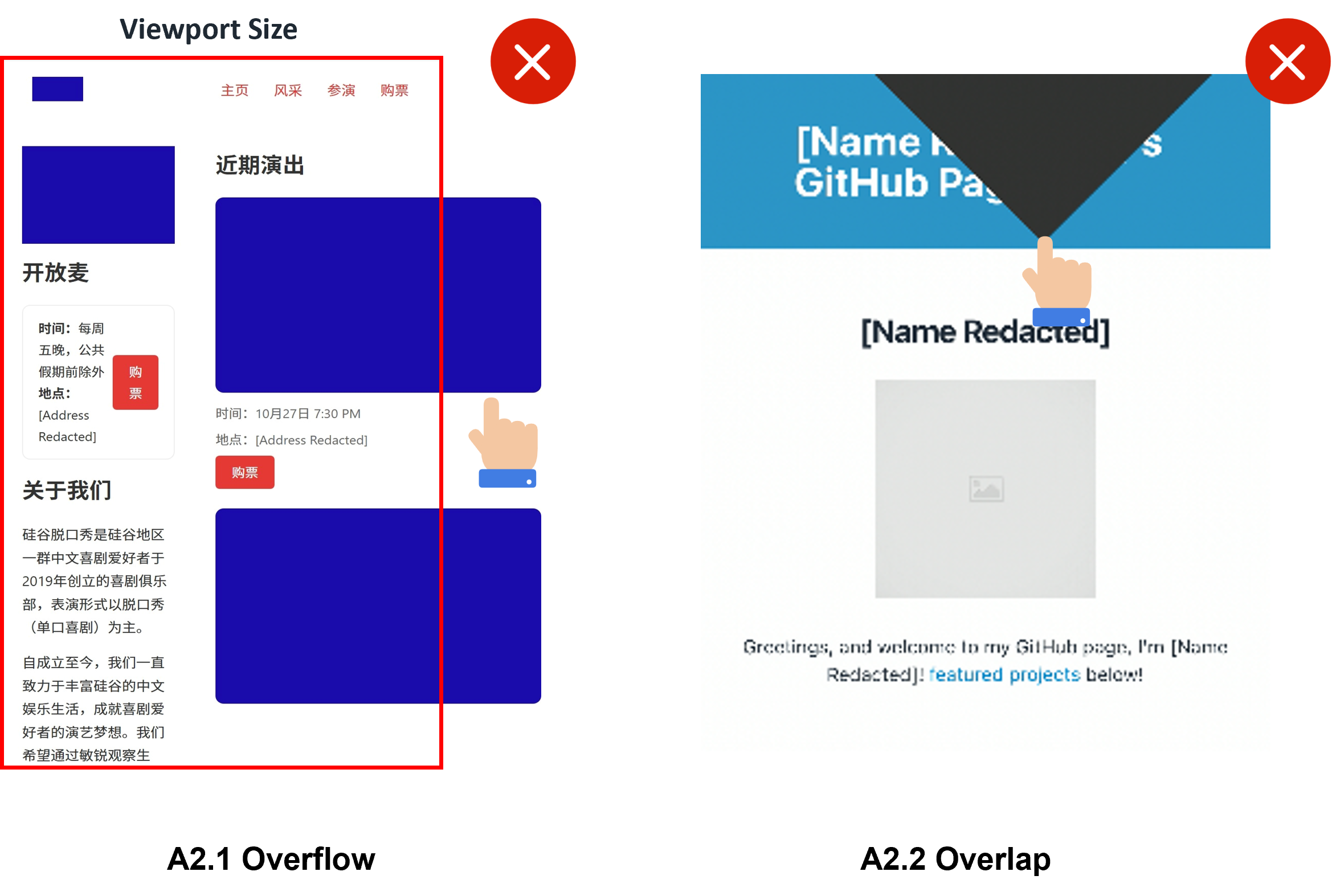}
  \caption{Subclass of A2 (Structural Layout Failure): A2.1 shows content overflow when elements exceed the viewport; A2.2 shows overlap caused by elements colliding after responsive reflow.}
  \label{fig:subclass-A2}
\end{figure}

\begin{figure}[t]
  \centering
  \includegraphics[width=\linewidth]{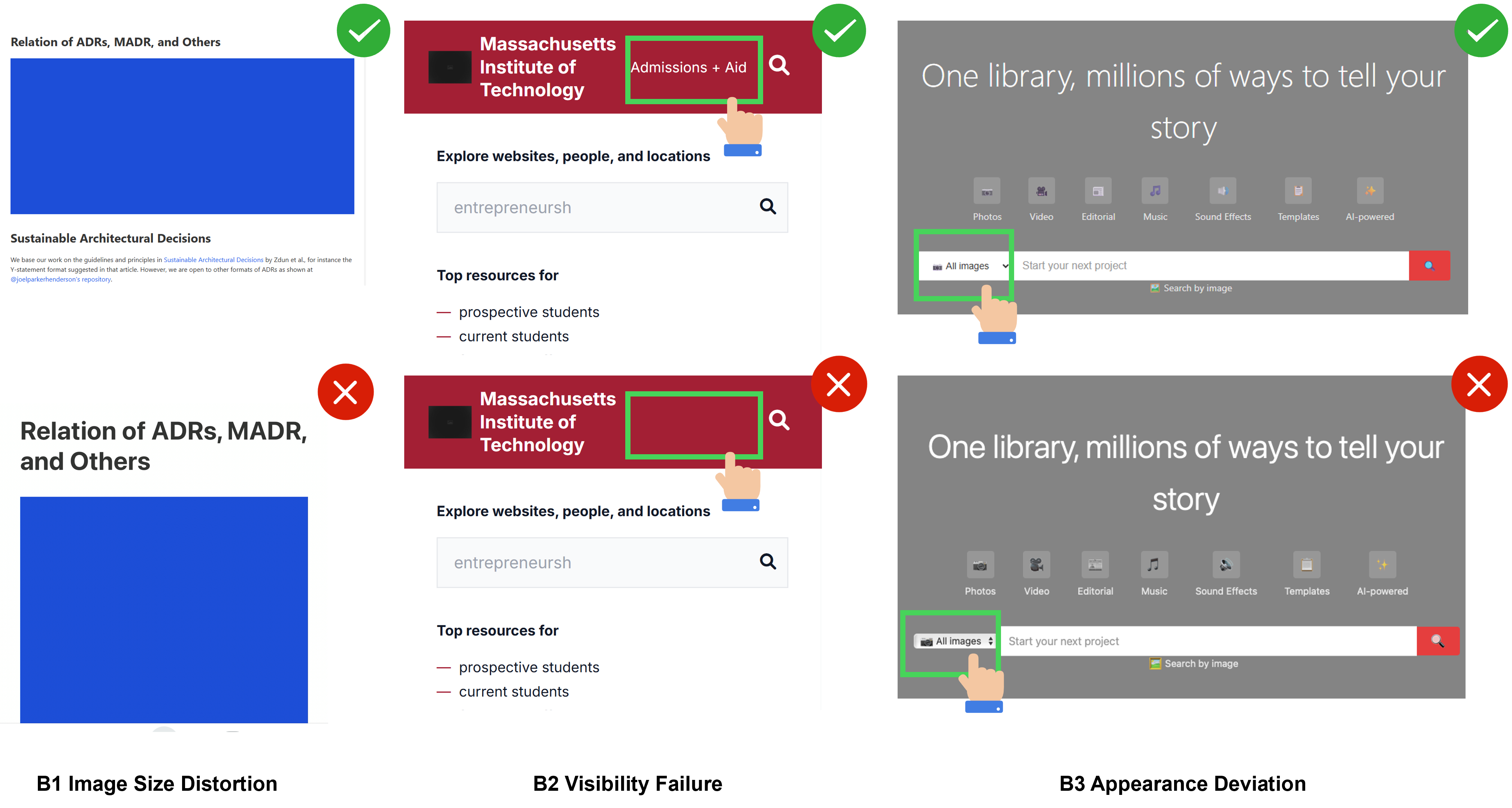}
  \caption{Examples of Class B element-level incompatibilities: B1 (image size distortion), B2 (visibility failure), and B3 (appearance deviation) in AI-generated webpages.}
  \label{fig:subclass-B}
\end{figure}

Page-level failures are further divided into two subtypes: \textit{viewport adaptation (A1)} and \textit{structural layout failure (A2)}. 
Viewport adaptation failure occurs when the page preserves desktop element spatial relationships despite being rendered on smaller viewports, causing cramped layouts and unreadable content (Fig.~\ref{fig:subclass-A1}). In contrast, structural layout failure covers cases where the page \emph{attempts} to restructure for mobile but produces broken layouts (Fig.~\ref{fig:subclass-A2}). 
Element-level failures (Class B) are localized to individual components: image size distortion (B1), visibility failure (B2), and appearance deviation (B3), as illustrated in Fig.~\ref{fig:subclass-B}.
In many cases, the generated webpages behave like desktop-oriented layouts and do not reflow according to the smaller viewport size of mobile devices.

\finding{Compatibility failure in AI-generated webpages are mostly page-level (88.3\%) with viewport adaptation issues.}


\subsection{RQ3: Root Causes of Compatibility Issue}

\textbf{Motivation.} Visible compatibility failures show how a page breaks. They do not explain \textit{why}, namely, which code patterns cause the breakage. The same viewport adaptation overflow failure may have different causes. Examples include a missing viewport declaration, fixed dimensions, and incomplete wrapping rules. We therefore examine the recurring HTML/CSS patterns behind these failures.

\textbf{Settings.} We analyze the 139 incompatible AI-generated webpages identified in RQ1. For each page, the same two annotators (\S\ref{sec:stage4}) inspect its HTML, CSS, and rendered screenshots, and independently assign one or more root-cause labels. We allow multiple labels because a page may contain several code defects, such as a missing viewport meta tag together with hard-coded widths. We measure agreement with per-category binary Cohen's kappa and obtain a macro-average kappa of 0.940 before resolving disagreements through discussion. We derive the categories from observed code patterns.

In addition, we conduct a keyword scan to examine the usage of responsive
layout constructs. Responsive constructs are CSS/HTML mechanisms that enable a webpage to adapt its layout to different screen sizes and rendering environments.
These include \texttt{flex}, \texttt{grid}, and Tailwind prefixes such as \texttt{sm:}, \texttt{md:}, and \texttt{lg:}. 
The scan provides context for interpreting whether failures occur despite the presence of
responsive layout constructs.

\textbf{Results.} Responsive constructs are widely-used in the generated code, but many pages still fail across rendering environments. In the current keyword scan, 97.5\% of instances contain at least one responsive layout construct. The keyword \texttt{flex} appears in 95.1\% of the instances, and the keyword \texttt{grid} appears in 43.3\%. At the same time, 139 generated pages in the empirical study set remain incompatible across environments. This contrast indicates that MLLMs are aware of responsive constructs in front-end coding, but  \textit{their implementation of these constructs is incomplete or incorrect}.
Table~\ref{tab:rq3-root-causes} reports six code-level root causes that turn responsive code into compatible failures. We detail them below.

\begin{table}[t]
\centering
\caption{Code-level root causes among incompatible AI-generated webpages.}
\label{tab:rq3-root-causes}
\small
\begin{tabular}{lrr}
\toprule
  \textbf{Root cause} & \textbf{\#Pages} & \textbf{Rate} \\
\midrule
Missing viewport meta & 64 & 46.0\% \\
Missing flexible wrapping & 41 & 29.5\% \\
Hard-coded dimensions & 27 & 19.4\% \\
Hard-coded image size & 14 & 10.1\% \\
Hidden element & 3 & 2.2\% \\
Browser rendering anomaly & 2 & 1.4\% \\
\bottomrule
\end{tabular}
\end{table}

\smallskip
\noindent\textit{Missing viewport meta} (46.0\%). 
Pages that omit or misuse \texttt{<meta name="viewport">} fail to specify intended layout width to mobile. So mobile browsers default to rendering the page at desktop viewport width, then scale it down to fit the physical screen. This shrinking causes text or elements to be too small, manifesting as symptoms A1.1 or A1.2.

\smallskip
\noindent\textit{Missing flexible wrapping} (29.5\%). Pages implement
\texttt{flex} or \texttt{grid} but omit \texttt{flex-wrap} or equivalent
wrapping rules. In such a case, containers force child elements to remain on a single line, even when the viewport is too narrow, mostly causing A2.

\smallskip
\noindent\textit{Hard-coded dimensions} (19.4\%). Fixed layout values lock element sizes (\texttt{min-width}, \texttt{width: ...px}). On smaller viewports, 
such values either exceed available space (causing overflow) or fail to scale responsively, leading to A1.2 or A2.1.

\smallskip
\noindent\textit{Hard-coded image size} (10.1\%). Responsive width utilities \texttt{w-full} adapt to viewport width, but pairing with fixed heights or aspect-ratio assumptions that don't scale. The image stretches disproportionately, causing B1.

\smallskip
\noindent\textit{Hidden element} (2.2\%). Responsive visibility utilities (\texttt{hidden lg:block}) hide elements on mobile without offering a replacement entry point. As a result, the hidden elements become unreachable in the mobile rendering, leading to B2.

\smallskip
\noindent\textit{Browser rendering anomaly} (1.4\%). Browser-specific defaults in styling, spacing, or native control rendering diverge across browsers and mainly lead to B3.

\finding{While AI-generated webpages often include responsive layout constructs, they fall short of correct implementation. The dominant causes are missing viewport meta (46\%) and missing flexible wrapping (29.5\%).}

\section{XCompat: Compatibility Detection Tool}
\label{sec:xcompat}

Our empirical study shows that compatibility failures in AI-generated webpages are common. To detect these failures before production, we propose an automated, lightweight compatibility issue detection tool, \textsc{XCompat}, based on post-rendered visual signals.
We formulate the detection task at the same pair granularity as in previous sections.

\subsection{XCompat Design}
\label{subsec:xcompat-impl}

For each reference-target pair, \textsc{XCompat} first extracts signals from the rendered DOM and screenshot, applies the two analysis modules, and then combines their module results into a single pair-level compatibility decision.
Specifically, it uses a Page-Level Analysis Module for global layout and viewport problems and an Element-Level Analysis Module for localized component problems. 

To start with, \textsc{XCompat} utilizes post-render HTML DOM tree, screenshots, and viewport metadata. This information will be fed into the two modules below.

\textit{Page-Level Analysis Module.}
This module mainly detects (Class A) page-level failures, including viewport adaptation and structural layout failures. Motivated by the finding that a large proportion of incompatibilities stem from missing or misconfigured viewport metadata, we start by inspecting code-level viewport configurations, including default width settings, target viewport-width, content fill ratio, and horizontal overflow, using heuristic predicate thresholds. 
For structural analysis, it identifies whitespace anomalies by detecting abnormal blank areas in the target screenshot but absent in the reference. It further detects layout collapse by computing overlaps among multiple elements in the rendered DOM tree. A substantial discrepancy in overlaps between rendering environments indicates conflicting global layout structures and is flagged as a compatibility failure.

\textit{Element-Level Analysis Module.}
This module targets localized compatibility failures on individual matched DOM elements across rendering environments (Class B). 
Taking per-element rendering signals such as bounding boxes, aspect ratios, and visibility state as inputs, the module detects two classes of failures motivated by our incompatibility taxonomy. First, it identifies distorted elements by comparing aspect ratios between matched elements; a substantial deviation indicates abnormal stretching or compression caused by environment-specific rendering. Second, it captures visibility inconsistencies by examining rendering-specific states such as \texttt{display} or \texttt{opacity}; discrepancies in these signals reveal elements that are unintentionally changed between renderings. 

\textit{Aggregation.}
The two above modules generate results at different granularities. \textsc{XCompat} finally combines them with a logical operator \textsc{or}. A pair is marked \textsc{Incompatible} if either module returns a compatibility failure.

\subsection{Experiment Setup}
\label{subsec:xcompat-setup}
\label{sec:xcompat-eval}

We evaluate \textsc{XCompat} on the following research questions:

\begin{itemize}[leftmargin=*,itemsep=2pt]
  \item \textbf{(RQ4)} \textit{How accurately does \textsc{XCompat} detect cross-environment incompatibilities?} We compare it with existing compatibility checking tool and LLM-based baselines.
  \item \textbf{(RQ5)} \textit{How efficient is \textsc{XCompat}?} We conduct the runtime and monetary cost for our tool.
  
\end{itemize}
\begin{table*}[t]
  \centering
  \caption{Incompatibility detection results on the empirical and test sets.}
  \label{tab:xcompat-effectiveness}
  \small
  \setlength{\tabcolsep}{5pt}
  \begin{tabular}{lrrrrrrrr}
    \toprule
    \multirow{2}{*}{\textbf{Method}} &
    \multicolumn{4}{c}{\textbf{Empirical Set} ($N=1{,}624$)} &
    \multicolumn{4}{c}{\textbf{Test Set} ($N=408$)} \\
    \cmidrule(lr){2-5}\cmidrule(lr){6-9}
    & \textbf{P} & \textbf{R} & \textbf{F1} & \textbf{Acc}
    & \textbf{P} & \textbf{R} & \textbf{F1} & \textbf{Acc} \\
    \midrule
    \multicolumn{9}{l}{\textit{Compatibility testing tools}} \\
    \textsc{RedeCheck} & 0.411 & 0.904 & 0.565 & 0.651 & 0.342 & 0.657 & 0.450 & 0.610 \\
    \midrule
    \multicolumn{9}{l}{\textit{LLM-based baselines}} \\
    GPT-5.5 DOM-only & 0.926 & 0.828 & 0.874 & 0.940 & 0.899 & 0.808 & 0.851 & 0.931 \\
    GPT-5.5 DOM w/ screenshot & 0.733 & 0.899 & 0.808 & 0.893 & 0.693 &\textbf{0.889} & 0.779 & 0.877 \\
    Opus 4.8 DOM-only & 0.926 & 0.860 & 0.892 & 0.948 & 0.909 & 0.808 & 0.856 & 0.934 \\
    Opus 4.8 DOM w/ screenshot & 0.828 & 0.710 & 0.765 & 0.890 & 0.788 & 0.677 & 0.728 & 0.877 \\
    \midrule
    \textsc{XCompat}  (Ours) & \textbf{0.932} &\textbf{0.946}  &\textbf{0.939}  &\textbf{0.969}  & \textbf{0.917} & \textbf{0.889} & \textbf{0.903} & \textbf{0.953} \\
    \bottomrule
    \bottomrule
    \multicolumn{9}{l}{\textit{Ablation study}} \\
    \textsc{XCompat} w/o Page-Level Module & 0.793 & 0.160 & 0.266 & 0.779 & 0.905 & 0.192 & 0.317 & 0.799 \\
    \textsc{XCompat} w/o Element-Level Module &0.964  & 0.848 & 0.902 & 0.954 &0.929  & 0.798 & 0.859 & 0.936 \\    
    \bottomrule
  \end{tabular}
\end{table*}

\subsubsection{Dataset}
Section~\ref{sec:dataset} is split 8:2 into an empirical set and a test set. The empirical set supports design intuitions, threshold selection, prompt selection for LLM baselines, and baseline sanity checks. To avoid data leakage, we fix all configurations and separately evaluate \textsc{XCompat} on the 20\% unseen test set. We also report the performance difference between the two sets.

\subsubsection{Baselines}
We compare \textsc{XCompat} with existing compatibility testing tools and two LLM-based baselines. We do not include \textsc{X-Pert} or WebDiff as primary baselines because they primarily target cross-browser inconsistencies at a fixed viewport, whereas our benchmark is dominated by cross-device responsive-layout issues, as shown in RQ1.
\textsc{RedeCheck}~\cite{redecheck} analyzes layout structure during viewport resizing and serves as a DOM-geometry-based compatibility baseline.
For LLM-based baselines, we evaluate two input settings for each model to assess whether visual rendering evidence improves compatibility issue detection. For both settings, we provide the same taxonomy definitions used in our annotation protocol. The \textit{DOM-only setting} receives post-render DOM snapshots, environment names, and the comparison mode. The \textit{DOM w/ screenshot setting} additionally receives the reference and target screenshots. We choose GPT-5.5 and Claude Opus 4.8 for comparison.

\subsubsection{Metrics}
Following prior work on cross-browser incompatibility and responsive-layout failure detection~\cite{choudhary2013x,redecheck,watanabe2023layout}, we report \textit{Precision, Recall, and F1} for incompatibility detection results. 
We additionally report \textit{Accuracy} since our detection task is evaluated as pair-level binary classification. 
For efficiency, we report \textit{execution time per comparison} and estimated \textit{monetary cost} per 1{,}000 comparisons when using online model providers.

\subsection{RQ4: Effectiveness of XCompat}
\label{subsec:rq-eff}

Table~\ref{tab:xcompat-effectiveness} reports detection results on both the empirical set and the test set. The empirical set is used for empirical analysis in Section~\ref{sec:empirical}, while the test set is entirely unseen, demonstrating \textsc{XCompat}'s effectiveness on unseen webpages.

\textsc{XCompat} achieves the highest F1 on both sets. 
On the test set, \textsc{XCompat} achieves Precision of 0.917, Recall of 0.889, F1 of 0.903, and Accuracy of 0.953 over 408 instances, meaning it can detect 91\% compatibility issues in AI-generated webpages. On the empirical set, \textsc{XCompat} reaches F1 of 0.939, compared with 0.892 for the strongest LLM baseline and 0.565 for \textsc{RedeCheck}. 
The modest F1 drop of 3.6\% from the empirical set to the unseen test set indicates that the fixed detector generalizes well on unseen pages and remains stronger than both structural and LLM-based baselines.

The ablation study demonstrates that both page-level and element-level modules effectively contribute to our method (F1 score). The greater performance drop without the page-level module corroborates our finding of dominant page-level issues. The page-level-only variant prioritizes precision but reduces recall and overall F1 performance due to increased false negatives.
We also conduct a sensitivity analysis for three manually specified predicate thresholds on the test set. Varying the viewport-width threshold from 800 px to 760 and 840 px, the content-fill threshold from 0.75 to 0.70 and 0.80, and the mobile wide-content threshold from 2.0 to 1.8 and 2.2 does not change any pair-level binary prediction.

In MLLM baselines, we observe that DOM-only inputs already provide useful evidence for compatibility classification. Adding screenshot inputs reduces performance for both models. GPT-5.5 drops from 0.851 to 0.779 F1 on the test set. Opus 4.8 drops from 0.856 to 0.728 F1 on the test set. A likely explanation is that full-page screenshots introduce high-dimensional visual noise and multiple artifacts (responsive changes, scaling effects, visually salient but harmless differences) that distract from clearer DOM signals on viewport size, overflow flags, and element geometry.

\subsection{RQ5: Efficiency and Cost}
\label{subsec:rq-eff2}

RQ5 evaluates whether \textsc{XCompat} is practical for repeated multi-environment compatibility testing by analyzing its execution costs. Table~\ref{tab:xcompat-efficiency} reports the execution time and cost on the test set, compared to baselines.

\begin{table}[t]
  \centering
  \caption{Runtime and token cost on the test set. Runtime is reported as mean execution time per comparison over 408 test-set pairs. Token cost reports mean total API tokens per logged LLM call. Offline tools do not consume API tokens.}
  \label{tab:xcompat-efficiency}
  \small
  \setlength{\tabcolsep}{5pt}
  \begin{tabular}{lrr}
    \toprule
    \textbf{Method} & \textbf{$\bar{t}$ / Comp.} & \textbf{Total Tok. / Call} \\
    \midrule
    \textsc{RedeCheck} & 0.068\,s & 0 \\
    GPT-5.5 DOM-only & 17.53\,s & 15{,}304 \\
    GPT-5.5 DOM w/ screenshot & 27.96\,s & 20{,}721 \\
    Opus 4.8 DOM-only & 6.49\,s & 22{,}050 \\
    Opus 4.8 DOM w/ screenshot & 8.45\,s & 26{,}721 \\
    \midrule
    \textsc{XCompat} & 0.127\,s & 0 \\
    \bottomrule
  \end{tabular}
\end{table}

\textsc{XCompat} runs as a deterministic offline analysis over pre-captured renderings, eliminating network dependencies and API token consumption. It processes each sample in 0.127 seconds, approximately 51 times faster than the fastest LLM baseline.
LLM baselines further incur 15{,}304 to 26{,}721 tokens per API call, corresponding to \$0.0934 to \$0.1407 per comparison. Although ReDeCheck also runs offline and requires 0.068 seconds per comparison, as shown in RQ4, this speed comes at the cost of substantially lower detection effectiveness.

\subsection{Failure Study}
\label{sec:failure-study}

We further analyzed \textsc{XCompat} errors on the test set. Among the 408 reference--target pairs, \textsc{XCompat} makes 19 errors, including 11 false negatives and 8 false positives.

The false negatives mostly correspond to subtle compatibility problems that are hard to detect automatically. 
Five cases involves content extending beyond the screen width, where our detector neglects evidence of overflow due to conservative thresholds.
Three cases visually appeares to contain a desktop layout without proper mobile adjustment,  where main content was narrow and centered, 
but it does not meet the overflow pattern in our detector.
The remaining false negatives include two image-size distortions and one whitespace anomaly, where the visual difference is too minor to be distinguished by the current DOM and screenshot signals. 
For the false positives, \textsc{XCompat} detects minor coordinate changes 
, but the changes are visually negligible under our annotation criteria.

Overall, these errors show that \textsc{XCompat} fails in two main situations. First, it may miss compatibility issues when the visual defects in DOM and screenshot are subtle. Second, it may over-report benign geometric changes that resemble compatibility failures but do not affect the perceived layout.

\section{Threats to Validity}
\label{sec:threats}

\textit{Source page selection.} 
Our empirical chooses self-contained and static webpages. This is because they provide deterministic comparison, where rendering behavior is reproducible across environments without depending on user interactions or backend state.
Nevertheless, the detector operates on rendered screenshots and post-render DOMs, making it technically applicable to any webpage rendering comparisons with the same UI state.
We note that responsive design principles apply across framework-based development (Vue, React).
While CSS and UI frameworks provide responsive assistants, the final behavior still depends on how framework classes and custom CSS are used. Therefore, our findings remain relevant to framework-based projects.

\textit{Prompt formulation.}
We used consistent, neutral prompts across all AI tools without explicit instructions for cross-environment compatibility, ensuring fair comparison without special prompt engineering effects. This reflects a baseline performance. Prompts emphasizing cross-environment compatibility could improve results but risk over-specialized layouts.

\textit{Annotation Subjectivity.}
A construct threat is the subjectivity of compatibility labeling. 
Cross-environment rendering naturally introduces visual differences, and some of them are acceptable responsive adaptations. 
We mitigate this threat by defining compatibility issues operationally as user-perceivable failures affecting visibility, layout, or readability. 
We further standardize the annotation process with a detailed guideline and pilot labeling involving two independent annotators. 
Disagreements are resolved through discussion, and we report Cohen's $\kappa$ to quantify inter-annotator agreement.

\textit{Missing design intents.} Another source of ambiguity is missing design intent. 
Some visual differences may be deliberate, such as image cropping for a full-bleed background or logo clipping within a constrained container. 
We treat responsive rearrangement as valid behavior and label size/appearance changes as failures only when they cause abnormal distortion or missing styles. Since borderline cases remain subjective, element-level findings should be interpreted conservatively.

\section{Conclusion and Future Work}

We presented the first systematic study of cross-environment compatibility in AI-generated webpages. 
Using \textsc{WebCompat}, we found that compatibility issues are widespread, with cross-device failures occurring far more frequently than pure cross-browser failures. 
Most issues affect page-level layout, readability, and viewport adaptation, and are associated with recurring implementation patterns such as hard-coded dimensions, missing viewport settings, unflexible wrapping, and rigid image sizing. 
Based on these findings, we developed \textsc{XCompat}, which combines screenshots and DOM snapshots to detect compatibility issues. It outperforms existing structural and multimodal LLM baselines. 
Our results demonstrate that single-environment visual fidelity is insufficient for evaluating AI-generated webpages. 
Future work will extend the benchmark to more interactive and production-oriented webpages and explore automated root cause localization and repair.

\section{Data Availability}
The artifact is available at: \url{https://github.com/ZiyunGuo/WebCompat}.

\balance
\bibliographystyle{IEEEtran}
\bibliography{software}

@inproceedings{mesbah2011automated,
  title={Automated cross-browser compatibility testing},
  author={Mesbah, Ali and Prasad, Mukul R},
  booktitle={Proceedings of the 33rd International Conference on Software Engineering},
  pages={561--570},
  year={2011}
}

@article{saar2016browserbite,
  title={Browserbite: cross-browser testing via image processing},
  author={Saar, T{\~o}nis and Dumas, Marlon and Kaljuve, Marti and Semenenko, Nataliia},
  journal={Software: Practice and Experience},
  volume={46},
  number={11},
  pages={1459--1477},
  year={2016},
  publisher={Wiley Online Library}
}

@article{wan2026runnable,
  title={From Runnable to Shippable: Multi-Agent Test-Driven Development for Generating Full-Stack Web Applications from Requirements},
  author={Wan, Yuxuan and Liang, Tingshuo and Xu, Jiakai and Xiao, Jingyu and Huo, Yintong and Lyu, Michael R},
  journal={arXiv preprint arXiv:2605.17242},
  year={2026}
}

@article{wu2025mllm,
  title={Mllm-based ui2code automation guided by ui layout information},
  author={Wu, Fan and Gao, Cuiyun and Li, Shuqing and Wen, Xin-Cheng and Liao, Qing},
  journal={Proceedings of the ACM on Software Engineering},
  volume={2},
  number={ISSTA},
  pages={1123--1145},
  year={2025},
  publisher={ACM New York, NY, USA}
}

@inproceedings{choudhary2010webdiff,
  title={WEBDIFF: Automated identification of cross-browser issues in web applications},
  author={Choudhary, Shauvik Roy and Versee, Husayn and Orso, Alessandro},
  booktitle={2010 IEEE International Conference on Software Maintenance},
  pages={1--10},
  year={2010},
  organization={IEEE}
}

@inproceedings{crosscheck,
  author    = {Shauvik Roy Choudhary and Mukul R. Prasad and Alessandro Orso},
  title     = {Combining Crawling and Differencing to Better
               Detect Cross-browser Incompatibilities in Web Applications},
  booktitle = {Proceedings of the 5th IEEE International Conference on
               Software Testing, Verification and Validation (ICST)},
  pages     = {171--180},
  year      = {2012},
  publisher = {IEEE},
  doi       = {10.1109/ICST.2012.97}
}

@inproceedings{redecheck,
  author    = {Thomas A. Walsh and Gregory M. Kapfhammer and Phil McMinn},
  title     = {Automated Layout Failure Detection for Responsive Web Pages
               without an Explicit Oracle},
  booktitle = {Proceedings of the 26th ACM SIGSOFT International Symposium
               on Software Testing and Analysis (ISSTA)},
  pages     = {192--202},
  year      = {2017},
  publisher = {ACM},
  doi       = {10.1145/3092703.3092712}
}

@inproceedings{choudhary2013x,
  title={X-PERT: Accurate identification of cross-browser issues in web applications},
  author={Choudhary, Shauvik Roy and Prasad, Mukul R and Orso, Alessandro},
  booktitle={2013 35th International Conference on Software Engineering (ICSE)},
  pages={702--711},
  year={2013},
  organization={IEEE}
}

@inproceedings{si2025design2code,
  title={Design2code: Benchmarking multimodal code generation for automated front-end engineering},
  author={Si, Chenglei and Zhang, Yanzhe and Li, Ryan and Yang, Zhengyuan and Liu, Ruibo and Yang, Diyi},
  booktitle={Proceedings of the 2025 Conference of the Nations of the Americas Chapter of the Association for Computational Linguistics: Human Language Technologies (Volume 1: Long Papers)},
  pages={3956--3974},
  year={2025}
}

@article{yun2024web2code,
  title={Web2code: A large-scale webpage-to-code dataset and evaluation framework for multimodal llms},
  author={Yun, Sukmin and Lin, Haokun and Thushara, Rusiru and Bhat, Mohammad Q and Wang, Yongxin and Jiang, Zutao and Deng, Mingkai and Wang, Jinhong and Tao, Tianhua and Li, Junbo and others},
  journal={Advances in neural information processing systems},
  volume={37},
  pages={112134--112157},
  year={2024}
}

@misc{googleGeminiCanvas,
  author = {{Google}},
  title = {{Gemini Canvas: Write, Code, and Create in One Space with AI}},
  howpublished = {[Online]. Available: \url{https://gemini.google/overview/canvas/}},
  note = {Accessed: Jun. 18, 2026}
}

@misc{vercelV0Screenshots,
  author = {{Vercel}},
  title = {{Screenshots and Files}},
  howpublished = {[Online]. Available: \url{https://v0.app/docs/screenshots}},
  note = {Accessed: Jun. 18, 2026}
}

@misc{cursorProduct,
  author = {{Cursor}},
  title = {{Build Software with AI Agents}},
  howpublished = {[Online]. Available: \url{https://cursor.com/product}},
  note = {Accessed: Jun. 18, 2026}
}

@inproceedings{xiao2024interaction2code,
author = {Xiao, Jingyu and Wan, Yuxuan and Huo, Yintong and Wang, Zixin and Xu, Xinyi and Wang, Wenxuan and Xu, Zhiyao and Wang, Yuhang and Lyu, Michael R.},
title = {Interaction2Code: Benchmarking MLLM-based Interactive Webpage Code Generation from Interactive Prototyping},
year = {2025},
publisher = {IEEE Press},
url = {https://doi.org/10.1109/ASE63991.2025.00028},
doi = {10.1109/ASE63991.2025.00028},
booktitle = {2025 40th IEEE/ACM International Conference on Automated Software Engineering (ASE)},
pages = {241–253},
numpages = {13},
location = {Seoul, Korea, Republic of}
}

@article{xiao2025designbench,
  title={Designbench: A comprehensive benchmark for mllm-based front-end code generation},
  author={Xiao, Jingyu and Wang, Ming and Lam, Man Ho and Wan, Yuxuan and Liu, Junliang and Huo, Yintong and Lyu, Michael R},
  journal={arXiv preprint arXiv:2506.06251},
  doi = {10.48550/arXiv.2506.06251},
  year={2025}
}

@inproceedings{liang2025waffle,
  title={WAFFLE: Fine-tuning Multi-Modal Model for Automated Front-End Development},
  author={Liang, Shanchao and Jiang, Nan and Qian, Shangshu and Tan, Lin},
  booktitle={Proceedings of the 63rd Annual Meeting of the Association for Computational Linguistics (Volume 1: Long Papers)},
  pages={24786--24802},
  year={2025}
}

@article{wan2025dcgen,
  title={Divide-and-conquer: Generating ui code from screenshots},
  author={Wan, Yuxuan and Wang, Chaozheng and Dong, Yi and Wang, Wenxuan and Li, Shuqing and Huo, Yintong and Lyu, Michael},
  journal={Proceedings of the ACM on Software Engineering},
  volume={2},
  number={FSE},
  pages={2099--2122},
  year={2025},
  publisher={ACM New York, NY, USA}
}

@inproceedings{gui2025uicopilot,
  title={Uicopilot: Automating ui synthesis via hierarchical code generation from webpage designs},
  author={Gui, Yi and Wan, Yao and Li, Zhen and Zhang, Zhongyi and Chen, Dongping and Zhang, Hongyu and Su, Yi and Chen, Bohua and Zhou, Xing and Jiang, Wenbin and others},
  booktitle={Proceedings of the ACM on Web Conference 2025},
  pages={1846--1855},
  year={2025}
}

@article{xiao2026comuicoder,
  title={Comuicoder: Component-based reusable ui code generation for complex websites via semantic segmentation and element-wise feedback},
  author={Xiao, Jingyu and Qin, Jiantong and Li, Shuoqi and Lam, Man Ho and Wan, Yuxuan and Huang, Jen-tse and Huo, Yintong and Lyu, Michael R},
  journal={arXiv preprint arXiv:2602.19276},
  year={2026}
}

@article{xiao2025efficientuicoder,
  title={Efficientuicoder: Efficient mllm-based ui code generation via input and output token compression},
  author={Xiao, Jingyu and Zhang, Zhongyi and Wan, Yuxuan and Huo, Yintong and Liu, Yang and Lyu, Michael R},
  journal={arXiv preprint arXiv:2509.12159},
  year={2025}
}

@misc{statcounter_browser_2026,
  title        = {Browser Market Share Worldwide},
  author       = {{StatCounter Global Stats}},
  howpublished = {\url{https://gs.statcounter.com/browser-market-share}},
  note         = {Accessed: 2026-06-28}
}

@misc{statcounter_desktop_os_2026,
  title        = {Desktop Operating System Market Share Worldwide},
  author       = {{StatCounter Global Stats}},
  howpublished = {\url{https://gs.statcounter.com/os-market-share/desktop/worldwide}},
  note         = {Accessed: 2026-06-28}
}

@article{cohen1960coefficient,
  title={A coefficient of agreement for nominal scales},
  author={Cohen, Jacob},
  journal={Educational and psychological measurement},
  volume={20},
  number={1},
  pages={37--46},
  year={1960},
  publisher={Sage Publications Sage CA: Thousand Oaks, CA}
}

@misc{browserstack2026,
  title        = {BrowserStack},
  author       = {{BrowserStack}},
  year         = {2026},
  howpublished = {\url{https://www.browserstack.com/}},
  note         = {Accessed: 2026-06-29}
}

@article{achiam2023gpt,
  title={Gpt-4 technical report},
  author={Achiam, Josh and Adler, Steven and Agarwal, Sandhini and Ahmad, Lama and Akkaya, Ilge and Aleman, Florencia Leoni and Almeida, Diogo and Altenschmidt, Janko and Altman, Sam and Anadkat, Shyamal and others},
  journal={arXiv preprint arXiv:2303.08774},
  year={2023}
}

@misc{openai2025gpt51,
  title        = {GPT-5.1: A Smarter, More Conversational ChatGPT},
  author       = {{OpenAI}},
  year         = {2025},
  howpublished = {\url{https://openai.com/index/gpt-5-1/}},
  note         = {Accessed: 2026-06-29}
}

@article{le2026uibenchkit,
  title={UIBenchKit: A unified toolkit for design-to-code model evaluation},
  author={Le, Chinh T and Siang, Trevor Ong Yee and Xiao, Jingyu and Wan, Yuxuan and Huo, Yintong},
  journal={arXiv preprint arXiv:2605.13141},
  year={2026}
}

@article{li2025webdevjudge,
  title={Webdevjudge: Evaluating (m) llms as critiques for web development quality},
  author={Li, Chunyang and Zheng, Yilun and Huang, Xinting and Fang, Tianqing and Xu, Jiahao and Chen, Lihui and Song, Yangqiu and Hu, Han},
  journal={arXiv preprint arXiv:2510.18560},
  year={2025}
}

@inproceedings{papineni2002bleu,
  title={Bleu: a method for automatic evaluation of machine translation},
  author={Papineni, Kishore and Roukos, Salim and Ward, Todd and Zhu, Wei-Jing},
  booktitle={Proceedings of the 40th annual meeting of the Association for Computational Linguistics},
  pages={311--318},
  year={2002}
}

@inproceedings{radford2021learning,
  title={Learning transferable visual models from natural language supervision},
  author={Radford, Alec and Kim, Jong Wook and Hallacy, Chris and Ramesh, Aditya and Goh, Gabriel and Agarwal, Sandhini and Sastry, Girish and Askell, Amanda and Mishkin, Pamela and Clark, Jack and others},
  booktitle={International conference on machine learning},
  pages={8748--8763},
  year={2021},
  organization={PmLR}
}

@article{wang2004image,
  title={Image quality assessment: from error visibility to structural similarity},
  author={Wang, Zhou and Bovik, Alan C and Sheikh, Hamid R and Simoncelli, Eero P},
  journal={IEEE transactions on image processing},
  volume={13},
  number={4},
  pages={600--612},
  year={2004},
  publisher={IEEE}
}

@article{watanabe2023layout,
  title={Layout cross-browser failure classification for mobile responsive design Web applications: Combining classification models using feature selection},
  author={Watanabe, Willian Massami and Dos Santos, Danilo Alves and De Oliveira, Claiton},
  journal={ACM Transactions on the Web},
  volume={17},
  number={4},
  pages={1--34},
  year={2023},
  publisher={ACM New York, NY}
}

@inproceedings{chen2018ui,
  title={From ui design image to gui skeleton: a neural machine translator to bootstrap mobile gui implementation},
  author={Chen, Chunyang and Su, Ting and Meng, Guozhu and Xing, Zhenchang and Liu, Yang},
  booktitle={Proceedings of the 40th International Conference on Software Engineering},
  pages={665--676},
  year={2018}
}

@article{chen2026designcoder,
  title={Designcoder: hierarchy-aware and self-correcting ui code generation with large language models},
  author={Chen, Yunnong and Yu, Xinyu and Ding, Shixian and Zhang, Yingying and Shi, Chengwei and Du, Jingzhou and Chen, Liuqing},
  journal={Information and Software Technology},
  pages={108214},
  year={2026},
  publisher={Elsevier}
}

@inproceedings{yuan2025designrepair,
  title={Designrepair: Dual-stream design guideline-aware frontend repair with large language models},
  author={Yuan, Mingyue and Chen, Jieshan and Xing, Zhenchang and Quigley, Aaron and Luo, Yuyu and Luo, Tianqi and Mohammadi, Gelareh and Lu, Qinghua and Zhu, Liming},
  booktitle={2025 IEEE/ACM 47th International Conference on Software Engineering (ICSE)},
  pages={2483--2494},
  year={2025},
  organization={IEEE}
}

@inproceedings{wei2016taming,
  title={Taming android fragmentation: Characterizing and detecting compatibility issues for android apps},
  author={Wei, Lili and Liu, Yepang and Cheung, Shing-Chi},
  booktitle={Proceedings of the 31st IEEE/ACM international conference on automated software engineering},
  pages={226--237},
  year={2016}
}

@inproceedings{huang2023conffix,
  title={Conffix: Repairing configuration compatibility issues in android apps},
  author={Huang, Huaxun and Xu, Chi and Wen, Ming and Liu, Yepang and Cheung, Shing-Chi},
  booktitle={Proceedings of the 32nd ACM SIGSOFT International Symposium on Software Testing and Analysis},
  pages={514--525},
  year={2023}
}

@inproceedings{liu2020owl,
  title={Owl eyes: Spotting ui display issues via visual understanding},
  author={Liu, Zhe and Chen, Chunyang and Wang, Junjie and Huang, Yuekai and Hu, Jun and Wang, Qing},
  booktitle={Proceedings of the 35th IEEE/ACM international conference on automated software engineering},
  pages={398--409},
  year={2020}
}

@inproceedings{zhao2020seenomaly,
  title={Seenomaly: Vision-based linting of gui animation effects against design-don't guidelines},
  author={Zhao, Dehai and Xing, Zhenchang and Chen, Chunyang and Xu, Xiwei and Zhu, Liming and Li, Guoqiang and Wang, Jinshui},
  booktitle={Proceedings of the ACM/IEEE 42nd international conference on software engineering},
  pages={1286--1297},
  year={2020}
}

\end{document}